\documentclass[reprint,amsmath,amssymb,prx]{revtex4-2} 

\usepackage{graphicx}
\usepackage{dcolumn}
\usepackage{bm}
\usepackage[normalem]{ulem}
\usepackage{comment}
\usepackage[dvipsnames]{xcolor}
\usepackage{float}
\usepackage{adjustbox}
\usepackage{lineno}
\usepackage{setspace}
\usepackage{natbib}
\usepackage{lipsum}
\usepackage[labeled]{multibib}

\newcommand{\ket}[1]{\left|#1\right\rangle}
\newcommand{\bra}[1]{\left\langle#1\right|}

\newcommand{\dif}{\mathrm{d}\;\!\!}
\newcommand{\matrixG}{\mathcal{G}}

\begin{document}

\title{Super- and subradiant dynamics of quantum emitters \\mediated by atomic matter waves
}

\author{Youngshin Kim}
\author{Alfonso Lanuza}
\author{Dominik Schneble}
\affiliation{Department of Physics and Astronomy, Stony Brook University, Stony Brook, NY 11794-3800, USA}

\date{\today}

\begin{abstract}

The cooperative modification of spontaneous radiative decay is a paradigmatic many-emitter effect in quantum optics. So far its experimental realization has involved interactions mediated by rapidly escaping photons that do not play an active role in the emitter dynamics. Here we explore cooperative dynamics of quantum emitters in an optical lattice that interact by radiating atomic matter waves. Using the ability to prepare weakly and strongly interacting many-body phases of excitations in an array of matter-wave emitters, we demonstrate directional super- and subradiance from a superfluid phase with tunable radiative phase lags, and directly access the buildup of coherence imprinted by the emitted radiation across a Mott insulator. We investigate the onset of cooperative dynamics for slow wave propagation and observe a coupling to collective bound states with radiation trapped at and between the emitters. Our results in open-system quantum electrodynamics establish ultracold matter waves as a versatile tool for studying many-body quantum optics in spatially extended and ordered systems.
\end{abstract}

\maketitle

\section*{Introduction}

Dicke's theory of super- and subradiance \cite{Dicke1954}, which describes how the spontaneous emission of photons can be enhanced or suppressed due to collective states of quantum emitters, introduced one of the earliest and central problems in quantum optics \cite{Rehler1971,Gross1982,Benedict1996,Brandes2005}. In the standard description invoking the Born-Markov approximation \cite{Gross1982,Tannoudji1998}, the photonic degrees of freedom are traced out, leaving only the system of quantum emitters with effective interactions between them. It is predicated on the assumption that the photons mediate such interactions instantaneously and that they are irreversibly lost to the vacuum. While this picture greatly reduces the complexity of problems and is broadly applicable in optical experiments with photons traveling through emitter ensembles within a fraction of the decay time \cite{Arecchi1970,Gross1982}, it tends to obscure the role that the photons play as the mediators of interactions in collective dynamics. This aspect becomes especially important for ensembles of distant emitters, with separations on the order of the wavelength, a subject that has recently seen a surge of theoretical interest \cite{Scully2006,Porras2008,Tudela2017prl,Asenjo2017,Zhang2019,Zhong2020,Sinha2020,Masson2022,Sheremet2023}.

Photon-mediated interactions in free space are generally weak between distant emitters \cite{deVoe1996}, and strong cooperative effects in extended ensembles can only be attained with special initial states or with a small number of available resonant modes per emitter \cite{Guerin2017}. While the spontaneous buildup of coherence across emitters leading to a burst of radiation \cite{Dicke1954,Rehler1971,Skribanowitz1973} requires sub-wavelength separation \cite{Masson2020}, superradiant states have been realized as single-excitation timed Dicke states \cite{Scully2006}, in which the phase lag of radiation from emitters at different positions results in constructive interference for a specific wavevector and highly-directional emission \cite{Araujo2016,Roof2016}. Subradiant states, which are harder to access due to their decoupling from the environment, have only been indirectly observed in atomic clouds with high cooperativity \cite{Guerin2016,Ferioli2021}.

The free-space limitations can be overcome via coupling to guided modes that support long-range interactions \cite{Goban2015,Solano2017,Corzo2019,Pennetta2022} in the paradigm of waveguide quantum electrodynamics \cite{Chang2018,Sheremet2023}. With emitters positioned at or in a one-dimensional waveguide, extended samples can undergo a buildup of coherence \cite{Liedl2023}, and super- and subradiant states can exist even in a small number of distant emitters \cite{vanLoo2013,Mirhosseini2019,Zanner2022,Kannan2023,Tiranov2023}. In this setting, the dynamics of photons can play an important role, as propagation delays or trapping of radiation between cavity-like geometries of emitters in waveguides can exert coherent backaction, defying a Markovian picture \cite{Zheng2013,Pichler2016,Sinha2020,Sinha2020a,Hsu2016,Chen2016,Leonforte2021}. Nonlinear dispersions near bandgaps can further modify the behavior of guided photons \cite{Bykov1975,Calajo2016,Tudela2017prl,Sanchez-Burillo2017,Chang2018} by slowing down or binding them to form evanescent fields around emitters near photonic crystals \cite{Goban2015,Chang2018}. While modifications of decay dynamics due to these non-Markovian properties were observed for single emitters \cite{Lodahl2004,Ferreira2021}, their effects on cooperative many-emitter dynamics remain largely unexplored.

In this paper, we investigate the role of radiation in the cooperative dynamics of an array of matter-wave emitters, using atoms in lieu of photons to mediate interactions between them. Our emitters, which are coupled to a one-dimensional vacuum featuring a band edge \cite{deVega2008,Navarrete2011,Krinner2018,Stewart2020}, are realized in a tunable optical lattice that allows us to initialize them in well-controlled many-body states, including single-excitation timed-Dicke states in a superfluid, and incoherent, fully inverted states in a Mott phase. For systematic studies of collective radiative behavior, the vacuum coupling and excitation energy of the emitters can be independently controlled.

We demonstrate directional super- and subradiance at variable normalized spacings $d/\lambda$ and phases $\phi$ between emitters, where $\lambda$ is the de Broglie wavelength of the emitted matter waves.
Their slow propagation, on the time scale of the decay, makes it possible to observe radiative dynamics before the radiation spreads through the array \cite{Lanuza2022}, giving access to a nascent phase of super- and subradiance.
Going beyond radiative retardation effects \cite{Arecchi1970,Sinha2020}, we uncover the role of collective emitter-photon bound states near the edge \cite{Bykov1975,Calajo2016,Sanchez-Burillo2017} and within \cite{Hsu2016,Chen2016,Leonforte2021} the continuum in a non-Markovian regime.
While the limited coherence length of matter waves in our system prevents a direct observation of a superradiant burst from a Mott phase \cite{deVega2008,Navarrete2011}, we probe the radiation-induced buildup of coherence  \cite{Dicke1954,Rehler1971} across the array by accessing its quasimomentum distribution, which has not been achieved in photon-based experiments. Finally, we explore a novel mechanism toward dissipative state engineering, in which a buildup of coherence is induced in a non-decaying atomic state acting as a spectator.

\begin{figure}[!htbp]
\centering
    \includegraphics[width=.99\columnwidth]{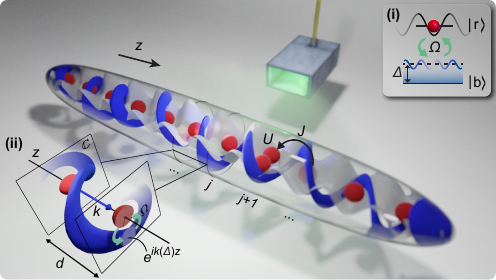}
    \caption{ \textbf{Schematics of the experimental setup.}
    A state-selective optical lattice (along $z$, in gray) traps atoms in state $\ket r$  (red) and emits matter-wave radiation in state $\ket {b}$ (blue) upon application of microwave coupling (green) between both states; atoms are transversely confined in lattice tubes of variable depth, with only one shown. The initial state is controlled via a Bose-Hubbard model with tunneling $J$ and on-site interactions $U$. In the example shown, the complex amplitude of the emitted matter-wave radiation twists along the spatial coordinate $z$, representing directional emission from a phase-coherent initial state (with parameters $\Delta=2.25\omega_\textsf{r}$, $\phi=-\pi/2$; see main text).
    Insets: (i) Coupling mechanism.
    On a given site $j$, the microwave field with strength $\Omega$ couples the trapped state, $\hat{r}_j^\dag \ket 0$, to a continuum of momentum states, $\hat{b}_k^\dag \ket 0$, around the resonance $k(\Delta)$.
    The detuning $\hbar\Delta$ from the $k=0$ state defines the excitation energy of the emitters, and determines the de Broglie wavelength $\lambda = 2\pi/k(\Delta)$ of the radiation. (ii) Illustration of the phase evolution of the radiation (with momentum $k(\Delta)=1.5k_\textsf{r}$).
    } 
\label{FIG1}
\end{figure}

\section*{Experimental scheme}

The experimental scheme is shown in Fig.~\ref{FIG1}. We implement a Bose-Hubbard model of $^{87}$Rb atoms in a ``red" state $\ket{r}=\ket{F=1,m_F=-1}$ in an optical lattice with variable depths $s_z (s_{\perp})$ in the longitudinal (transverse) directions, measured in units of the respective recoil energy $E_{\textsf{r},i} =\hbar^2/2 m\lambda_i^2$ where $\lambda_z(\lambda_\perp)=790(1064)\mbox{nm}$ and $m$ is the atomic mass. A resonant microwave field at $6.8~\mbox{GHz}$ couples the trapped $\ket{r}$ atoms to a second, ``blue" state $\ket{b}=\ket{2,0}$, which is made insensitive to the $z$ lattice by an appropriate choice of its wavelength and polarization, and is thus free to move longitudinally (i.e. along tubes defined by $s_\perp$, on time scales shorter than the residual confinement $2\pi\omega_z^{-1}\sim10~\mbox{ms}$; Methods).

Applying the external microwave field converts the $z$ lattice with its sites $j$  into an array of matter-wave emitters. The coherent coupling of the atomic excitations $\hat{r}^\dag_j \ket0$  to the continuum of momentum states $\hat{b}_k^\dag \ket0$ provides a mechanism for spontaneous decay \cite{deVega2008,Krinner2018} (cf. Fig.~\ref{FIG1} (i)), whose properties are controlled by the strength $\Omega$ and detuning $\Delta$ of the microwave coupling from the edge at $k=0$. Upon loss from the lattice, the excitation energy $\hbar\Delta$ is converted into kinetic energy  $\hbar \omega_k = \hbar^2 k^2/2m$ of matter waves centered at the resonant momenta $k(\Delta) = \pm \sqrt{\Delta/\omega_\textsf{r}}k_\textsf{r}$ and de Broglie wavelength $\lambda(\Delta) = 2\pi/k(\Delta)$, where $k_\textsf{r}=2\pi/\lambda_z$ and $\omega_\textsf{r} = E_{\textsf{r},z}/\hbar = 2\pi \times 3.7$~kHz. The dynamics are modeled by a sum of Weisskopf-Wigner-type couplings \cite{Stewart2017,Lanuza2022} $H' = (\hbar\Omega/2) \sum_{j,k} \gamma_{k,j}\hat{b}_k^\dag\hat{r}_j + \text{H.c.}$ with the free Hamiltonian $\hat{H}_0 = \sum_j \hbar\Delta \hat{r}^\dag_j \hat{r}_j + \sum_k \hbar\omega_k \hat{b}^\dag_k \hat{b}_k$. The wavefunction overlap (Franck-Condon factor) $\gamma_{k,j} = \bra{0} \hat{b}_k \hat{r}_j^\dag \ket{0}$ encodes the phase lag $e^{-ikdj}$ for emission from different emitters (sites) spaced by $d=\lambda_z/2=\pi/k_\textsf{r}$. The character of the initial state in the emitter array is adjusted between coherent (superfluid) and fully inverted (Mott insulating) by adiabatically controlling the ratio of the on-site interactions $U$ and the tunneling rate $J$ ($\lesssim 2\pi \times 0.1$~kHz) before emission, via the lattice depths (Methods).

\begin{figure}[!htbp]
\centering
    \includegraphics[width=0.99\columnwidth]{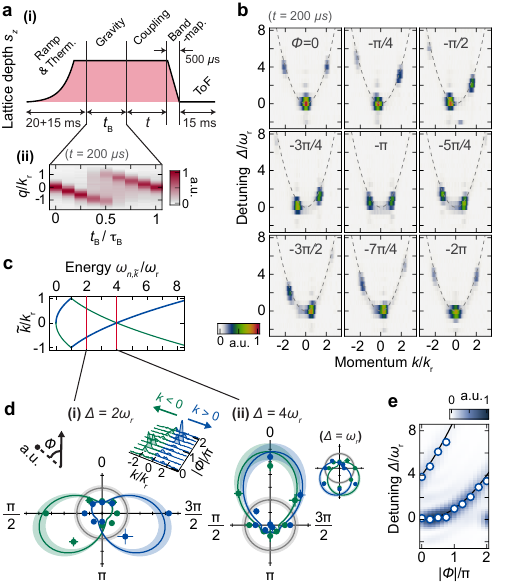}
    \caption{
    \textbf{Phase control of emission characteristics}. \textbf{a.} (i) Experimental sequence. A superfluid is prepared in the optical lattice $(s_z,s_\perp)=(8,0)$, and a differential phase $\phi$ between emitters is gravitationally imprinted for a duration $t_{\text{B}}$.
    Afterward, a $200$-$\mu$s-long coupling pulse with strength $\Omega/\omega_\textsf{r} = 0.60(4)$ is applied to initiate the emission. (ii) Quasimomentum $q$ distribution of the $\ket{r}$ atoms as a function of $t_{\text{B}}$.
    \textbf{b.} Momentum distributions of emitted matter waves versus $\Delta$, plotted for various $\phi$ (each panel is normalized independently). The dashed lines are the dispersion $\omega_k = \omega_\textsf{r} (k/k_\textsf{r})^2$ of the matter wave. 
    \textbf{c.} Dispersion of matter waves in the reduced zone scheme, $\omega_{n,\tilde{k}}$, generated by mapping $k$ to the first Brillouin zone, $\tilde{k} \in (-k_\textsf{r},k_\textsf{r}]$ (blue and green lines).
    \textbf{d.} Emitted population in the positive ($k>0$) and negative ($k<0$) $z$ directions (blue and green points) as a function of $\phi$ for (i) $\Delta = 2\omega_\textsf{r}$ and (ii) $\Delta = 4\omega_\textsf{r}$ (also for $\Delta = \omega_\textsf{r}$; right-most inset). The blue and green lines represent finite-size simulations with 4 sites, and the gray line is the Markovian single-emitter prediction (Methods; shaded areas represent the uncertainty in $\Omega$). Also shown is a 3D density plot of momentum distribution for $\Delta = 2\omega_\textsf{r}$ (middle inset).
    \textbf{e.} Simulated emitted population with $k < 0$ as a function of $\Delta$ and $\phi$, overlaid with the fitted centers of the experimental emission peaks (circles). The black lines are the peak positions predicted from the dispersion $\omega_{n,\tilde{k}}$ with $\tilde{k} = q(\phi)$ and $|n|=0,1$.
    All data are averages of at least 3 measurements; the error bars show the standard error of the mean.
    }
\label{FIG2}
\end{figure}

\section*{Directional collective emission}

In the superfluid regime, each $\ket{r}$ atom is in a coherent superposition $\propto\sum_j \hat{r}_j^\dag \ket{0}$ of Wannier states over lattice sites $j$, which generalizes to $\hat{r}_q^\dag \ket{0} \propto \sum_j e^{iqdj} \hat{r}_j^\dag \ket{0}$ for finite quasimomentum $q \in (-k_\textsf{r},k_\textsf{r}]$. Upon radiative coupling, this Bloch state becomes a single-excitation collective state (also known as a timed-Dicke state in free space \cite{Scully2006}; TDS in short), with a phase lag $\phi = q d$ for radiation spontaneously emitted from neighboring sites. In our experiment, $q$ can be varied by subjecting the $\ket r$ atoms in the lattice to a potential gradient. For this purpose, we temporarily switch off the confining optical trap for a variable duration $t_{\text{B}}$ to induce a Bloch oscillation (driven by gravity $g$) to $q =-2k_\textsf{r} t_{\text{B}}/\tau_{\text{B}}$ with the period $\tau_{\text{B}} = 2\pi\hbar/(mgd) = 1.2$~ms, before turning on the vacuum coupling (see Fig. \ref{FIG2}a). Data for emission from the TDS with variable $\phi$ are shown in Fig.~\ref{FIG2}b; as we vary the excitation energy $\hbar \Delta$ (or the effective array spacing $d/\lambda\leq\sqrt{2}$), the recorded momentum distributions are strongly peaked along the parabolic dispersion.

The collective effects on the spontaneous emission are fixed by the transition matrix element $\bra{0} \hat{b}_{k} \hat{H}' \hat{r}_q^\dag \ket{0}$ of the coupling Hamiltonian (Methods). It involves a sum over amplitudes with phase factors $\exp[{i(q-k(\Delta))dj}]$ on each site $j$, and thus the emitted matter waves can interfere constructively (or destructively) when $q - k(\Delta)= (2\pi/d)n$ with an integer $n$ (or a half-integer). By introducing a quasimomentum for matter waves, $\tilde{k} \in (-k_\textsf{r},k_\textsf{r}]$ where $k = \tilde{k} + 2nk_\textsf{r}$ with an integer $n$, the condition for constructive interference simplifies to $q=\tilde{k}(\Delta)$, cf. Fig. \ref{FIG2}c. As observed in Fig. \ref{FIG2}b and d, this implies that at a given excitation energy $\hbar\Delta$, the emission can be strongly enhanced or suppressed (such as for $\phi=0$ or $-\pi$ at $\Delta=4\omega_\textsf{r}$ in Fig. \ref{FIG2}d (ii)), or even directionally \cite{Scully2006} enhanced and suppressed (such as for $\phi=-\pi/2$ at $\Delta \approx 2\omega_\textsf{r}$ in Fig. \ref{FIG2}d (i), the case also represented in Fig. \ref{FIG1}).
We note that, because of the quadratic dispersion $\omega_{n,\tilde{k}} = \omega_\textsf{r}(\tilde{k}+2nk_\textsf{r})^2/k_\textsf{r}^2$ (in the reduced zone scheme; cf. Fig. \ref{FIG2}c), the sensitivity of the superradiant detuning $\Delta(\phi)$ to variations of $\phi$ increases with $n$ as shown in Fig. \ref{FIG2}e, which may be of interest for metrological applications. For $\phi=0$, the smallest non-zero detunings at which the emitted matter waves interfere constructively (destructively) are  $\Delta = 4\omega_\textsf{r} ~(\omega_\textsf{r})$, which is the situation that we consider in the following.

\begin{figure*}[!htbp]
\centering
   \includegraphics[width=0.85\textwidth]{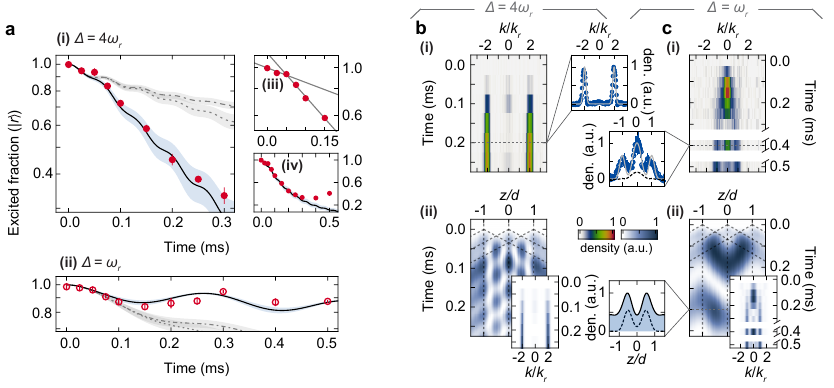}
    \caption{ \textbf{Super- and subradiant dynamics in the superfluid regime.} \textbf{a}, Excited fraction as a function of time for $(s_z,s_\perp)=(8,10)$, $\phi=0$ 
    at super- and subradiant detunings (i) $\Delta = 4\omega_\textsf{r}$ and (ii) $\Delta = \omega_\textsf{r}$ (red points and circles) with coupling strength adjusted to $\Omega/\omega_\textsf{r} = 1.00(7)$ and $\Omega/\omega_\textsf{r} = 0.42(3)$, respectively.
    The black solid (dot-dashed) lines are our simulations of an excitation spread over an array of three sites (localized in the middle site); the dotted line is that of an isolated site; shaded areas represent the uncertainty in $\Omega$.
    Also shown are excited fractions at (iii) early times and (iv) late times for $\Delta=4\omega_\textsf{r}$. The gray solid lines in (iii) are a piecewise linear fit, yielding the decay rates $\Gamma_{t<t_c}/\Gamma_1 = 1.1(2)$ and $\Gamma_{t>t_c}/\Gamma_1 = 3.2(2)$ with $t_c = 52(1)$~$\mu$s.
    \textbf{b}, (i) Momentum distribution of matter waves versus time for $\Delta = 4\omega_\textsf{r}$. The lineout plot shows the data at $0.2$~ms (blue points) along with our simulation (gray solid line). (ii) Simulated position and momentum distributions of matter waves versus time (Methods). The dashed diagonal and vertical lines are the `light-cones' $z(t) = dj \pm v_g t$ and the positions of the emitters.
    \textbf{c}, Same as \textbf{b}, but for $\Delta = \omega_\textsf{r}$. The lineouts are shown at $0.4$~ms (with data in circles and simulations in solid lines);
    also shown are the BIC contributions from our analytic theory as black dashed lines (Methods).
    All data are averages of at least $3$ measurements; the error bars show the standard error of the mean.
    }
\label{FIG3}
\end{figure*}

\section*{Collective dynamics}

We probe the radiative TDS dynamics at $\phi=0$ for moderately weak couplings $(\Omega/\Delta)^2 \ll 1$. 
Fig. \ref{FIG3}a shows the corresponding time evolutions, with $\Omega$ adjusted to give the same single-emitter decay rate $\Gamma_1 = 2\pi \times 0.24$~kHz (Methods).
For $\Delta=4\omega_\textsf{r}$ we observe a clear superradiant enhancement of the decay rate, which however does not appear immediately.
The idea of strict, instantaneous Markovianity would require the coherence length of the radiation to exceed the spatial extent of coherence in the emitter array \cite{Arecchi1970,Sinha2020}, which in our system is determined by the superfluid coherence length $\xi$ \cite{Gadway2012}. With an enhanced Markovian decay rate  $\Gamma_\xi$, this translates to $v_g/\Gamma_\xi \gg \xi$, a condition that for our small group velocities $v_g = \hbar |k(\Delta)|/m \sim 1 \mu\text{m}/\text{ms}$ is not strictly fulfilled.
It is thus necessary to consider the full non-Markovian dynamics, for which collective effects are expected to set in only once the emitted radiation reaches a neighboring site \cite{Sinha2020}.
From a piecewise exponential fit, we find that the decay rate changes from $\Gamma_1$ to $\Gamma_\xi$ after a time comparable to the propagation time $d/v_g = 34$~$\mu$s; its value $\Gamma_\xi \approx 3\Gamma_1$ is consistent with the predictions of a finite-size simulation and exact non-Markovian analysis with an effective number of emitters (3) as the only fitting parameter (Fig. \ref{FIGChi} and Methods). The fact that the effective array size is small compared to the size of our superfluid is consistent with phase fluctuations induced by interactions in the confining optical lattice \cite{Gadway2012}.
At late times we observe partial reabsorption, likely due to coupling to empty sites beyond the coherence length $\xi$ \cite{Lanuza2022}.

The momentum space distribution of the emitted radiation, shown in Fig. \ref{FIG3}b (i), provides a complementary picture of the dynamics. While the emission is peaked at $k= \pm 2k_\textsf{r}$, there is a small off-resonant contribution at zero momentum due to coupling to the continuum edge, which causes residual non-Markovian oscillations in the time evolution and is characteristic of single-emitter dynamics \cite{Lanuza2022}. The quantitative consistency of our model with the data motivates us to extrapolate to position space, cf. Fig. \ref{FIG3}b (ii). We see that the time scale for the observed onset of superradiance is consistent with the overlap of `light-cones' of matter waves emanating from neighboring sites \cite{Sinha2020}, which subsequently form standing waves.

For $\Delta=\omega_\textsf{r}$, the TDS exhibits suppressed oscillatory decay of the excited population (Fig. \ref{FIG3}a (ii)) that deviates from single-emitter behavior after the propagation time $d/v_g = 68$~$\mu$s. The momentum-space distribution, cf. Fig. \ref{FIG3}c (i), is dominated by an oscillation at zero momentum and smaller contributions at $\pm k_\textsf{r}$. Translated to position space, the population is mostly concentrated between the sites, a behavior reminiscent of a bound state in the continuum (BIC) \cite{Chen2016,Leonforte2021} as a polaritonic state \cite{Lanuza2022,Kwon2022} with radiation partially trapped in the cavity-like geometry formed by the emitters \cite{Leonforte2021}. A detailed analysis reveals that the oscillatory dynamics is dominated by a beating between a BIC with energy $1.02 \hbar\omega_\textsf{r}$ and a bound state at the continuum edge, with the population average approaching $\approx 0.83$ at long times (Fig. \ref{FIGPoles} and Methods).

Close to the continuum edge $\Delta \sim 0$, the radiative dynamics generally becomes strongly coupled. In this regime, array effects have been found to enhance non-Markovian oscillations of a single excited emitter surrounded by empty neighbors \cite{Krinner2018,Stewart2020}. For an initial TDS at $\Delta=0$, we observe an amplification of such array effects with an almost full amplitude and weak damping (Fig. \ref{FIGD0} and Methods). The observed behavior reflects the convergence, in the limit of diverging coherence length, of our TDSs dynamics toward a beat between polaritonic bands \cite{Reeves2015,Lanuza2022}.

\begin{figure*}[!htbp]
\centering
    \includegraphics[width=0.85\textwidth]{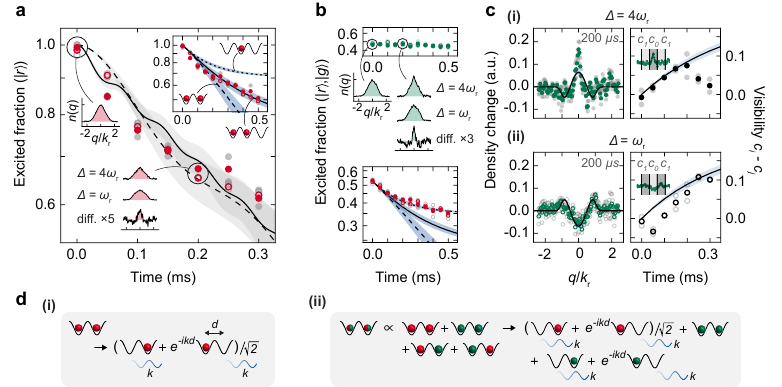}
    \caption{ \textbf{Spontaneous coherence formation in the Mott insulating regime.} \textbf{a,} Decay dynamics for $(s_z,s_\perp)=(15,40)$
    at detunings $\Delta = 4\omega_\textsf{r}$ and $\Delta = \omega_\textsf{r}$ (red points and circles) with coupling strengths adjusted to $\Omega/\omega_\textsf{r} = 1.00(7)$ and $\Omega/\omega_\textsf{r} = 0.50(4)$, respectively.
    The solid and dashed lines are our simulations of decay from an isolated site for the two detunings. Quasimomentum distributions (q.d.) of $\ket{r}$ atoms are shown for $t=0$ and $200$~$\mu$s. The top-right inset shows the same data including longer times, with master-equation simulations (Methods) of a 2-site model with and without an adjacent empty site (solid and dashed lines), and of a 1-site model with two adjacent empty sites \cite{Krinner2018} (dotted line); the model predictions coincide for both detunings.
    \textbf{b,} Decay of an equal superposition of the $\ket r$ and $\ket g$ (``red'' and ``green'') states in each site, with all other parameters as in {\bf a}. The top panel shows the fraction of $\ket{g}$ atoms (green circles and points), and corresponding q.d. at $t=0$ and $200$~$\mu$s. The bottom panel shows the fraction of $\ket{r}$ atoms (red circles and points) along with the predictions of master-equation simulations (scaled by a factor of $1.05$ to match the initial fraction). 
    \textbf{c,} (i) Change in the normalized q.d. for $\ket g$ atoms after $200$~$\mu$s (left), and visibility $c_0-c_1$ versus time (right) for $\Delta = 4\omega_\textsf{r}$; $c_0$ and $c_1$ are the integrations over $|q| \in [0,0.5k_\textsf{r}]$ and $|q| \in (0.5k_\textsf{r},1.5k_\textsf{r}]$, respectively. The solid lines are simulations of the 2-site model with an adjacent empty site. (ii) Same as in (i) but for $\Delta =\omega_\textsf{r}$, and showing visibility $c_1-c_0$.
    \textbf{d,} Mechanism for coherence formation. (i) Starting with $\ket r$ atoms in each site, emitting into the mode $k$ generates a relative phase factor $e^{-ikd}$ at the second site. (ii) Starting with a superposition of $\ket r$ and $\ket g$, the $\ket g$ atoms acquire coherence as the $\ket r$ atoms decay.
    The shaded areas around the curves in {\bf a}-{\bf c} indicate the uncertainty in $\Omega$. All data are averages of two measurements; raw data are shown in gray circles and points.
    }
\label{FIG4}
\end{figure*}

\section*{Coherence formation}

We now explore the Mott regime in a deep lattice in which the $\ket{r}$ atoms are localized in individual lattice sites. As before, we apply the vacuum coupling at $\Delta = 4\omega_\textsf{r}$ and $\omega_\textsf{r}$, with $\Omega$ adjusted to give the same Markovian single-emitter decay rate. As shown in Fig. \ref{FIG4}a, we find that the decays for both detunings slow down after starting out like those of isolated emitters. This is at variance with the expectation \cite{deVega2008} that the Mott phase is analogous to an array of excited emitters that decay in an accelerating superradiant cascade. However, the coherence length of a free gas of bosonic atoms in our deep one-dimensional tubes, i.e. that of the emitted radiation, is limited to $\sim d$ \cite{Gadway2012}, and we therefore expect that the effective number of emitters coherently coupled through the radiation is small. We consider arrays of several emitters (see Fig. \ref{FIG4}a inset) using a master-equation simulation (Methods), and find that the observed dynamics is best reproduced by an array of two populated sites, with a third, empty site representing the emitters outside the initially populated region, which can reabsorb the radiation \cite{Lanuza2022}.

In addition to the decay, the model predicts a buildup of coherence, as illustrated in Fig. \ref{FIG4}d, which can be understood by considering two emitters in the fully excited state $\hat{r}^\dag_0 \hat{r}^\dag_1 \ket{0}$. Each site radiates into the common mode at $k=k(\Delta)$ with a position-dependent phase, resulting in the state $(\hat{r}^\dag_1 + e^{-ikd} \hat{r}^\dag_0)\hat{b}^\dag_k\ket{0}/\sqrt{2}$ after the first emission, cf. Fig. \ref{FIG4}d (i). Extrapolated to a full lattice, the $\ket r$ part is a quasimomentum eigenstate, such that emission at $\Delta$ will lead to an enhancement around the quasimomentum $q=\tilde{k}(\Delta)$, i.e. $q=0(\pm k_\textsf{r})$ for $\Delta=4\omega_\textsf{r}(\omega_\textsf{r})$  (see also the dispersion relation in Fig. \ref{FIG2}c). Indeed, we observe a small difference in the shape of the quasimomentum distributions (q.d.) as the decay proceeds, revealing an excess of population near $q=0$ when comparing $\Delta=4\omega_\textsf{r}$ to $\Delta=\omega_\textsf{r}$ (see Fig. \ref{FIG4}a insets). Upon renormalizing them, such a change should also be visible with respect to the initial q.d., but we find that the feature is overshadowed by a broadening of both distributions, which is likely due to contaminations from higher lattice bands (Methods). To make the comparison possible, we introduce an auxiliary, radiatively-decoupled, ``green'' state $\ket{g}=\ket{2,1}$ in the $z$ lattice. We prepare this state, which experiences the same lattice potential and interactions as $\ket{r}$, in a coherent superposition with $\ket{r}$ on each site via a $\pi/2$ Rabi pulse. As illustrated in Fig. \ref{FIG4}d (ii), the $\ket{g}$ state can serve as a witness to the build-up of coherence induced by the decay.

Fig. \ref{FIG4}b shows the time evolution of the $\ket r$ and $\ket g$ populations for the same coupling parameters as before. The $\ket r$ atoms are now more sparse and decay more slowly, with the observed population dynamics following a scenario of a filled site surrounded by two empty sites (for the on-site interactions $U/\hbar\lesssim\Omega$, occupation by a $\ket{g}$ atom does not seem to prevent reabsorption). 
While the total $\ket g$ population remains unchanged, its q.d. exhibits marked differences depending on $\Delta$, similarly to those of $\ket{r}$. Since now there is no loss of population, the evolved q.d. can be directly compared to the q.d. at $t=0$. As shown in Fig. \ref{FIG4}c there is a growth of population at $q=0$ for $\Delta=4\omega_\textsf{r}$, and at $q=\pm k_\textsf{r}$ for $\Delta=\omega_\textsf{r}$, corresponding to the quasimomenta $\tilde{k}$ of the emitted matter waves. These features clearly appear as a result of coherence buildup in the emission process and may thus be considered as the precursor of a superradiant burst \cite{Dicke1954}. Their visibility is well described by our simple numerical model up to $\sim 0.2$~ms; the deviation at later times is consistent with dephasing on time scales $\hbar/U \approx (0.46 \omega_\textsf{r})^{-1}\sim 0.1$~ms corresponding to onsite interactions in the Mott phase. We note that without strong correlations, one would expect losses (instead of gains) at the emitted quasimomenta, $q = \tilde{k}(\Delta)$, a behavior that is indeed seen for the case of a thermal ensemble of $\ket{r}$ atoms (Fig. \ref{FIGTh} and Methods). This stark observed difference further illustrates that the coherence formation in the Mott phase is fundamentally a quantum effect that arises through collective dissipation.

\section*{Conclusion}

In this work, we have realized an array of distant quantum emitters interacting via atomic matter-wave radiation, which we found to play an integral role in determining collective sub- and superradiant emitter dynamics.
The radiative retardation, responsible for some of the novel effects reported here, depends on the effective separation $\eta = d / (v_g/\Gamma_1)$ \cite{Sinha2020} and may be made stronger by exploiting the scaling behaviors $v_g \propto \sqrt{\Delta}$ and $\Gamma_1 \propto\Omega^2/\sqrt{\Delta}$ in sufficiently deep lattices while maintaining weak coupling $(\Omega/\Delta)^2 \ll 1$. Several exotic phenomena, attributed to time-delayed quantum feedback, have been predicted beyond the regime $\eta \sim 0.1$ of this work, including enhancement or suppression exceeding standard Dicke super- and subradiance \cite{Dinc2019,Sinha2020}, Fano-like resonances in the radiation spectrum \cite{Sinha2020a}, and entanglement generation processes that are of interest for quantum networks \cite{Zheng2013,Pichler2016}. Extensions to higher-dimensional geometries or more complex vacuum structures \cite{Stewart2020,Lanuza2022} may give access to collective anisotropic emission \cite{Tudela2017prl} and subradiant self-guided modes \cite{Asenjo2017,Zhang2019}. Finally, our platform offers several unique opportunities at the intersection of quantum optics and condensed matter physics. Auxiliary atomic states, for example, may create novel possibilities for the dissipative engineering of quantum states \cite{Tudela2015}, while adding on-site optical disorder can allow access to localized phases \cite{Fayard2021} and their interplay with subradiant dynamics. 
Moreover, using strongly confining tubes, it should be possible to study the effects of fermionization of the radiation field, pointing towards fermionic quantum optics with matter waves \cite{Windt2023}.

\setcounter{figure}{0}
\renewcommand{\thefigure}{E\arabic{figure}}

\section*{Methods}

\subsection*{\label{APP:System} System preparation}

Each measurement begins with a Bose-Einstein condensate of around $2 \times 10^4$ ${}^{87}$Rb atoms in the $S_{1/2}$ hyperfine state $\ket r \equiv \ket{F=1, m_F = -1}$ (the ``red'' state) that is initially prepared in a nearly isotropic crossed optical-dipole trap (XODT) made of horizontal beams at a wavelength $\lambda_\perp = 1064$~nm. The transverse lattices of variable depth $s_\perp$ are implemented by partially retro-reflecting the XODT beams. The combined potential results in the residual harmonic confinement in the vertical $z$ direction with the trap frequency $\omega_z \in 2\pi \times [73,90]$~Hz for the range $s_\perp \in [0,40]$. The state-selective $z$ lattice of depth $s_z$ is made of $\sigma^-$-polarized vertical beams at a wavelength $\lambda_z = 790.0$~nm, which is tuned-out for the second hyperfine state $\ket{b} \equiv \ket{F=2, m_F=0}$ (the ``blue'' state) \cite{Krinner2018}. All the measurements are conducted in a constant bias field of $B_z = 5$~G in the $z$ direction to provide the quantization axis and the Zeeman splitting. 

The initial states of the $\ket r$ atoms in the optical lattices are prepared with an adiabatic exponential ramp of $s_\perp$ and $s_z$ over $\sim 200$~ms, and their many-body phases are determined by the tunneling rate $J$ (along $z$) and on-site interactions $U$. From our measurements of momentum peak width versus $s_\perp$ at a fixed $s_z = 15$, we estimate that the superfluid to Mott insulator transition occurs around $s_\perp=12$, at which the ratio $U/J$ is $\sim 40$ (cf. \cite{Greiner2002}). For the final depths $(s_z,s_\perp)=(8,10)$ and $(s_z,s_\perp)=(15,40)$ used to prepare a superfluid and a Mott insulator, our calculations yield $U/J = 6.5$ and $U/J = 84.5$, respectively. For these calculations, we use the band structures $\varepsilon_q(s_z)$ to obtain the (nearest-neighbor) tunneling rate along $z$, $J= -1/(2k_\textsf{r}) \int_{-k_\textsf{r}}^{k_\textsf{r}} dq e^{iqd} \varepsilon_q$, while Gaussian approximations of the Wannier functions $w(\vec{r})$ are used to compute the on-site interactions, $U = (4\pi \hbar^2 a/m) \int d\vec{r} |w(\vec{r})|^4$, where $a \approx 100a_0$ is the scattering length in terms of the Bohr radius $a_0$. For the preparation of superfluids with a nonzero phase ($q\neq0$) in Fig.~\ref{FIG2}, we momentarily turn off the XODT for a duration $t_{\text{B}}$, which induces a Bloch oscillation due to the gravitational potential. 

An externally applied microwave field of $6.8$~GHz provides the coupling between the $\ket r$ atoms in the $z$ lattice and the $\ket {b}$ atoms in one-dimensional free space. The coupling strength $\Omega$ is pre-calibrated via Rabi oscillation measurements; we use the peak-to-peak fluctuation of $7$~\% independently estimated from Rabi spectra taken over one month for the quoted uncertainty. The excitation energy of quantum emitters is given by the detuning $\Delta = \omega_\mu - \omega_{\text{res}}$ of the microwave field from the resonance frequency $\omega_{\text{res}} = \omega_{br} - \omega_{0}$ that includes the bare resonance $\omega_{br} (> 0)$ between the two hyperfine states, $\ket b$ and $\ket r$, and the zero-point energy $\omega_0$ of the $\ket r$ state in the $z$ lattice. We calibrate $\omega_{\text{res}}$ between each set of measurements (with the data for each figure divided into several such sets) via lattice transfer spectroscopy starting from a condensate in the $\ket b$ state \cite{Krinner2018}. 
For each figure, the measured values of $\omega_{\text{res}}$ fluctuated by $0.3$~kHz r.m.s, which is the dominant uncertainty in $\Delta$.

In Fig.~\ref{FIG4}, for the purpose of characterizing the coherence formation, we introduce another hyperfine ground state $\ket{g} = \ket{2,1}$ (the ``green'' state) without resonant coupling to the matter waves. This state experiences the same lattice potential as the $\ket{r} = \ket{1,-1}$ state, as its Land\'e factor $g_F$ only differs by a sign \cite{Grimm2000}. To generate the local superposition of the $\ket{g} = \ket{2,1}$ and $\ket{r} = \ket{1,-1}$ states, we prepare a Mott insulator of $\ket{r}$ atoms and apply a $1.3$-ms-long two-photon $\pi/2$ Rabi pulse made of a $6.8$-GHz microwave field and a $3.3$-MHz radio-frequency field to create an equal superposition of the two states in each site, $\propto \prod_j (\hat{r}_j^\dag + \hat{g}_j^\dag) \ket{0}$.

\subsection*{\label{APP:Imaging} Imaging and post-processing}

The detection sequence starts with a $500$-$\mu$s-long bandmapping step \cite{Stewart2020}, during which all lattices are ramped down to convert the quasimomenta into momenta. The XODT is then completely turned off to let the atoms expand freely in time of flight (ToF) lasting $12$~ms ($15$~ms) for the $\ket{F=2}$ ($\ket{F=1}$) state. During ToF, a magnetic field gradient is briefly applied to spatially separate the atoms in different Zeeman sublevels (Stern-Gerlach separation). For absorption imaging, light on the $D_2$ cycling transition $F=2$ $\rightarrow$ $F' = 3$  is first applied for $200$~$\mu$s to detect the $\ket{F=2}$ atoms, and after an additional ToF of $2.7$~ms, repump light on the $D_2$ transition $F=1$ $\rightarrow$ $F' = 2$ for $100$~$\mu$s and a second $200$-$\mu$s-long imaging pulse are applied to detect the $\ket{F=1}$ atoms. Additional empty pictures without atoms are collected to eliminate fringes in the images via principal component analysis. During ToF expansion, the $\ket{1,-1}$ atoms are partially redistributed to the $\ket{2,1}$ and $\ket{2,2}$ states with Rabi pulses to infer the magnetic field {\it in situ} \cite{Krinner2018a}. These separate images of $\ket r$ atoms transferred to different Zeeman states are later recombined to recover the original distributions of $\ket{r}$ atoms, which involves interpolating each image and rescaling the coordinates to compensate for different ToF times.

Various considerations are taken into account in analyzing the images. Due to residual heating during the ramping of lattices, a small fraction of $\ket r$ atoms are excited to higher bands that are decoupled from the matter waves. When counting the excited population, we only include the atoms in the ground band, $q \in [-k_\textsf{r},k_\textsf{r}]$ (with an additional tolerance of $\sigma_k \approx 0.15k_\textsf{r}$ accounting for the measurement resolution). In Fig.~\ref{FIG4}a, the fraction of $\ket{r}$ atoms outside the first Brillouin zone is on average $13$\% at early times $t \le 0.05$~ms but increases to $17$\% at later times $t \ge 0.2$~ms, effectively broadening the quasimomentum distribution over the decay time. To accurately determine the populations in each state in Fig.~\ref{FIG3} and~\ref{FIG4}, we subtract small background contributions contained in the $t=0$ images and additional empty images. The momentum coordinates in the images are calibrated via Kapitza-Dirac diffraction of the $\ket{1,-1}$ and $\ket{2,1}$ atoms \cite{Gadway2009}. During the $t_{\text{bm}}=500$~$\mu$s of bandmapping, the atoms are already moving, which contributes to the final distances traveled. For the $\ket{b}$ atoms, this time is added to the effective ToF time for the calibration. For the $\ket{g}$ and $\ket{r}$ atoms, the group velocity changes during the ramp down of the $z$ lattice. We estimate an effective correction to the TOF time by $1/(2k_\textsf{r})\int_{-k_\textsf{r}}^{k_\textsf{r}}dq \int_0^{t_{\text{bm}}} dt v_g(t)/v_g(t_{\text{bm}})$, where the instantaneous group velocity $v_g(t) = |d\omega_q/dq|$ depends on the value of $s_z$ at each time, yielding $0.3 t_{\text{bm}}$ ($0.5t_{\text{bm}}$) for $s_z = 15$ ($s_z = 8)$.

\subsection*{\label{APP:Fermi} Decay rates from Fermi's golden rule}

The transition rate of an initial discrete state $\ket i$ into a final state $\ket f$ in a continuum by a weak perturbation $\hat{H}'$ can be obtained from the Fermi's golden rule, $\Gamma_{i\rightarrow f} = (2\pi/\hbar) |\hat{H}_{fi}'|^2 \rho_{\text{dos}}(E_f=E_i)$, which depends on the transition matrix elements $H_{fi}' \equiv \bra{f} \hat{H}'\ket{i}$ and the density of the states $\rho_{\text{dos}}(E)$ \cite{Tannoudji1998}. Our system consists of the atomic excitations in the lattice, matter waves, and their couplings

\begin{align}
\label{Eq:H}
&\hat{H} = \sum_j \hbar\Delta \hat{r}^\dag_j \hat{r}_j + \sum_k \hbar\omega_k \hat{b}^\dag_k \hat{b}_k + H',
\\
&\text{where }H'=
\frac{\hbar\Omega}{2} \sum_{j,k} (\gamma_{j,k} \hat{r}_j^\dag \hat{b}_k +\gamma_{k,j}\hat{b}_k^\dag\hat{r}_j).\nonumber
\end{align}
The Franck-Condon factor is defined as the state overlap $\gamma_{j,k} =\gamma_{k,j}^*= \bra 0 \hat{r}_j \hat{b}_k^\dag \ket 0$. We approximate the Wannier states on individual lattice sites as Gaussians $\bra z \hat{r}_j^\dag \ket 0 \approx (\pi a_{\text{ho}}^2)^{-1/4} e^{-(z-dj)^2/(2a_{\text{ho}}^2)}$ with the harmonic oscillator length $a_{\text{ho}} = \sqrt{\hbar / m \omega_{\text{ho}}}$ and frequency $\omega_{\text{ho}} = 2\omega_\textsf{r} \sqrt{s_z}$, while the matter waves are taken to be free particles in a box of length $L$, $\bra z \hat{b}_k^\dag \ket 0 = L^{-1/2} e^{ikz}$. The effective system size $L$ can be taken to infinity as the experimental time scale is much shorter than that of the residual confinement. Taking the limit $L \rightarrow \infty$ yields $\gamma_{j,k} = \int_{-L/2}^{L/2} dz \bra 0 \hat{r}_j \ket z \bra z \hat{b}_k^\dag \ket 0 \approx \sqrt{2/L} (\pi a_{\text{ho}}^2)^{1/4} e^{-k^2 a_{\text{ho}}^2/2} e^{ikdj}$. Note that the dependence on $1/\sqrt{L}$ will be compensated during the continuum approximation of the summation $\sum_k \rightarrow (L/2\pi) \int dk$. One can also choose to normalize the matter waves over a Wigner-Seitz cell, e.g. $(-d/2,d/2]$, in which case $L$ is replaced by $d$ in $\gamma_{j,k}$.

We now calculate the decay rate for an initial state of $N$ atoms populating a timed-Dicke state in an array of $M$ sites, $\ket i = N!^{-1/2}\big(\hat{r}_{q,M}^\dag\big)^N \ket 0$ and a final state  $\ket f = (N-1)!^{-1/2}\hat{b}_k^\dag (\hat{r}_{q,M}^\dag)^{N-1} \ket 0$, where $\hat{r}_{q,M}^\dag = {M}^{-1/2} \sum_{j=\lfloor1-M/2\rfloor}^{\lfloor M/2 \rfloor} e^{iqdj} \hat{r}_j^\dag$ and the number of sites $M$ reflects the finite coherence length $\xi$ of the system. We assume that the bath of radiation modes is always empty, in accordance with the Markovian approximation. With the density of states of the matter waves $\rho_{\text{dos}}(E) = (L/2\pi\hbar) \sqrt{2m/E}$, the decay rate of the timed-Dicke state with quasimomentum $q$ and energy $\hbar\Delta$ follows as

\begin{align}
\Gamma_{q,M,N} &= (2\pi/\hbar) |\bra{i} \hat{H}' \ket f|^2 \rho_{\text{dos}}(\hbar\omega_k = \hbar\Delta)\\\nonumber
&= NM\Gamma_1 \left| \frac{1}{M} \sum_{j=\lfloor1-M/2\rfloor}^{\lfloor M/2 \rfloor}e^{i(q-k(\Delta))dj} \right|^2,
\end{align}
where $\Gamma_1 \equiv \Gamma_{q,M=1,N=1}$ is the single-emitter decay rate, $\Gamma_1 = (\Omega^2/\sqrt{\Delta})\sqrt{\pi/2\omega_{\text{ho}}} e^{-2\Delta/\omega_{\text{ho}}}$ \cite{Stewart2017,Lanuza2022}. The last factor leads to the modulation of the decay rate as a function of initial phase $\phi = qd$, which peaks at $q \equiv k(\Delta) ~{(\operatorname{mod } 2\pi/d)}$ and reduces to the Kronecker delta $\delta_{q\equiv k(\Delta)~{(\operatorname{mod } 2\pi/d)}}$ in the limit $M \rightarrow L/d$, consistent with the emission peaks in Fig.~\ref{FIG2}b.

\begin{figure}[!htbp]
\centering
    \includegraphics[width=0.99\columnwidth]{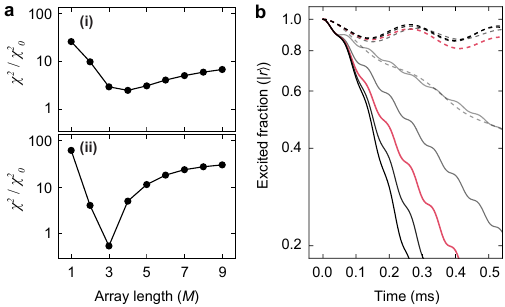}
    \caption{\textbf{Numerical simulations of an $M$-site array containing a single excitation.} \textbf{a,} $\chi^2$ vs. $M$ evaluated with respect to the data shown in Fig.~\ref{FIG2}d and Fig.~\ref{FIG3}a. In (i), the data at the phases $\phi \in (0, 2\pi]$ and the coupling parameters $\Delta/\omega_\textsf{r} = \{2,4\}$ are used (with $\Omega/\omega_\textsf{r} = 0.6$ and $t = 200$~$\mu$s). In (ii), the data at the coupling times $t \in [0,0.25]$~ms for $(\Delta,\Omega)/\omega_\textsf{r} = (4,1)$ and at $t \in [0,0.5]$~ms for $(\Delta,\Omega)/\omega_\textsf{r} = (1,0.42)$ are used. The normalization $\chi_0^2$ is the value at which the cumulative $\chi^2$ distribution reaches 95\%. The lines are guides for the eyes. \textbf{b,} Simulated population versus time for $(\Delta,\Omega)/\omega_\textsf{r} = (4,1)$ and $(1,0.42)$ (with $s_z=8$, $q=0$) shown as solid and dashed lines, with $M$ varying from one to five (lightest to darkest colors; red for $M=3$, the case also shown in Fig.~\ref{FIG3}a).
    }
\label{FIGChi}
\end{figure}

\subsection*{\label{APP:Numerics} Numerical simulation}

As described in the main text, we use a single excitation in a finite-size array model to describe the radiative dynamics at early times in the superfluid regimes. In a system of length $L$, the continuum of modes $k$ is discretized as $k_m = (2\pi/L)m$ with integer $m$. The matrix elements of the Hamiltonian (Eq. (\ref{Eq:H})) in the single-excitation manifold are given by $\bra{0} \hat{r}_j  \hat{H}  \hat{r}^\dag_{j'} \ket{0} = \hbar\Delta \delta_{jj'}$, $\bra{0} \hat{b}_{k_m}  \hat{H}  \hat{b}^\dag_{k_{m'}} \ket{0} = \hbar \omega_{k_m} \delta_{mm'}$, $\bra{0} \hat{r}_j \hat{H} \hat{b}^\dag_{k_m} \ket{0} = (\hbar\Omega/2) \gamma_{j,k_m}$, and its Hermitian conjugate. The Schr\"odinger equation of the discretized Hamiltonian can be solved with a cutoff $|k| \le k_\Lambda$ that is sufficiently high, i.e. $(k_\Lambda/k_\textsf{r})^2 \gg \Omega/\omega_\textsf{r},\Delta/\omega_\textsf{r}$, at the same time maintaining the system size large enough $L \gg v_g/\Gamma_1$ such that the boundaries have no effect. Starting from an initial timed-Dicke state coherently populating an array of $M$ sites with quasimomentum $q$, the excitation and matter-wave amplitudes can be obtained as a function of time as $\ket{\Psi(t)} = \sum_j A_j(t) \hat{r}_j^\dag \ket{0} + \sum_m B_{k_m}(t) \hat{b}_{k_m}^\dag \ket{0}$. We use $M$ as the only free parameter of the model reflecting the initial coherence length of the system, and determine $M$ from the best fit to the data. The $\chi^2$ values of the models with different $M$ are obtained by computing the differences between the observed ($P_{i}$) and calculated ($\tilde{P}_{i}$) excited fractions, i.e. $\chi^2 = \sum_i (P_{i} - \tilde{P}_{i})^2/\sigma_i^2$, where $\sigma_i$ is the experimental uncertainty (see Fig. \ref{FIGChi}a). The analysis yields the best-fit values of $M=4$ and $3$ for data in Fig.~\ref{FIG2}d and~\ref{FIG3}a, respectively. We use the amplitudes $B_{k_m}$ to generate the momentum distributions in Fig.~\ref{FIG2}e and Fig.~\ref{FIG3}b and c ($B_{z} = \sum_{k_m} L^{-1/2}e^{i k_m z} B_{k_m}$ for position space), with an additional Gaussian convolution of width $0.15 k_\textsf{r}$ to reflect the measurement resolution \cite{Stewart2020}, cf. Fig.~\ref{FIG3}b and c.

In order to simulate the decay dynamics in the strongly interacting regime in Fig.~\ref{FIG4}, we use a master equation for the reduced density matrix \cite{Tannoudji1998}, $\bar{\rho} = \text{Tr}_b \rho$, where the tracing is over the matter-wave modes, $\hat{b}_{k}^\dag \ket 0$. The master equation is given by the Lindblad form 
\begin{align}
\dot{\bar{\rho}} = [\tilde{H}, \bar{\rho}]/i\hbar + \sum_{jj'} (\Gamma_{jj'}/2)\left(2 \hat r_{j'} \bar{\rho} \hat r_j^\dag - \{\hat r_j^\dag \hat r_{j'}, \bar{\rho}\}\right), 
\end{align}
where $\tilde{H}/\hbar = \sum_j \Delta \hat r_j^\dag \hat r_j + \sum_j \Delta_g \hat g_j^\dag \hat g_j + \sum_{jj'} J_{jj'} \hat r_j^\dag \hat r_{j'}$. The effect of on-site interactions is included by allowing no more than a single excitation per site, e.g. $\hat{r}_j^\dag \hat{r}_j^\dag \ket 0 = 0$ (hardcore-boson limit).  The dissipative and coherent couplings are given as $\Gamma_{jj'}/2 + i J_{jj'} \equiv \hbar^{-2} \int^\infty_0 d\tau G_{j,j'}(\tau) e^{i\Delta \tau}$, where the bath correlation function is defined as $G_{j,j'}(\tau) = \sum_k (\hbar\Omega/2)^2 |\gamma_{0,k}|^2 e^{ikd(j-j')}e^{-i\omega_k\tau}$. The integrals are computed by applying the Sokhotski-Plemelj theorem, $\int_0^\infty d\tau (\dots) e^{i(\Delta-\omega_k)\tau} = (\dots) (\pi \delta (\Delta-\omega_k) + i \mathcal{P}(\Delta - \omega_k)^{-1})$ and carrying out the $k$ integral numerically. Once the density matrix $\bar{\rho}(t)$ is obtained, the excited population and quasimomentum distributions can be computed as $N_c = \text{Tr}[\bar{\rho} \sum_j \hat{c}_j^\dag\hat{c}_j]$ and $n_c(q) = \text{Tr} [\bar{\rho} \hat{c}^\dag_q \hat{c}_q]$, where $c \in \{r,g\}$ (for $\ket r$ and $\ket g$ atoms), again with an additional Gaussian convolution of width $0.15 k_\textsf{r}$.

\begin{table}[!htbp]
\centering
\begin{adjustbox}{max width=0.99\columnwidth}
\begin{tabular}{|c|c|c|c|}
 \multicolumn{4}{c}{ $\Delta =4\omega_r$  }\\\hline
   $p$ & \text{BS} & \text{sR} & \text{SR} \\\hline
 \multicolumn{1}{|c|}{$\omega_p$}   & $-0.019(4)$ & $4.07(1)-7(3) \times 10^{-6} i$  & $4.09(1)-0.10(1)i$\\
\multicolumn{1}{|c|}{ $\vec{A}_{p,0}$}
 &
  $\left(\begin{array}{c}
     0.0038(7) \\
     0.0043(8) \\
     0.0038(7)
 \end{array}\right) $
 & 
 $ \left(\begin{array}{c}
 -0.000034(9)-0.0036(5) i \\
 -0.00020(5)+0.0071(9) i \\
-0.000034(9)-0.0036(5) i
  \end{array}\right) $
 &
   $\left(\begin{array}{c}
0.611(4)+0.042(8) i \\
0.597(2)+0.027(5) i \\
0.611(4)+0.042(8) i 
  \end{array}\right) $
\\\hline
 \multicolumn{4}{c}{ $\Delta =\omega_r$  }\\\hline
   $p$ & \text{BS} & \text{sR} & \text{SR} \\\hline
 \multicolumn{1}{|c|}{$\omega_p$}  & $-0.010(2)$ & $1.023(3)-3(1) \times 10^{-6} i$  & $1.041(9)-0.11(2)i$  
 \\
 \multicolumn{1}{|c|}{ $\vec{A}_{p,0}$}
 &
  $ \left(\begin{array}{c}
0.009(1) \\
0.010(2) \\
0.009(1)
  \end{array}\right) $
 &
  $ \left(\begin{array}{c}
0.371(2)-0.0039(5) i \\
0.742(4)+0.0013(2) i \\
0.371(2)-0.0039(5) i 
  \end{array}\right) $
 &
  $ \left(\begin{array}{c}
0.237(5)+0.044(2) i \\
-0.225(3)-0.036(9) i \\
0.237(5)+0.044(2) i 
  \end{array}\right)$\\\hline

 \multicolumn{4}{c}{ $\Delta =0$  }\\\hline
   $p$ & \text{BS}$_1$ & \text{BS}$_2$ & \text{SR} \\\hline
 \multicolumn{1}{|c|}{$\omega_p$}  & $-0.22(2)$ & $-0.10(1)$  & $0.28(2) - 0.033(1) i$  
 \\
 \multicolumn{1}{|c|}{ $\vec{A}_{p,0}$}
 &
  $ \left(\begin{array}{c}
0.281(1) \\
0.36749(1) \\
0.281(1)
  \end{array}\right) $
 &
  $ \left(\begin{array}{c}
0.042(1)\\
-0.067(1) \\
0.042(1)
  \end{array}\right) $
 &
  $ \left(\begin{array}{c}
0.214(4)+0.100(4) i \\
0.3460(7) + 0.0137(6) i \\
0.214(4)+0.100(4) i 
  \end{array}\right)$\\\hline
\end{tabular}
\end{adjustbox}
\caption{Complex frequencies $\omega_p$ (in units of $\omega_\textsf{r}$) and initial amplitudes $\vec{A}_{p,0}$ of the bound (BS), subradiant (sR), and superradiant (SR) states impacting the dynamics of 3 quantum emitters, for the parameters tested in the experiment (see Figs.~\ref{FIG3} and \ref{FIGD0}).}
\label{TABLE1}
\end{table}

\begin{figure*}[!htbp]
\centering
    \includegraphics[width=.99\textwidth]{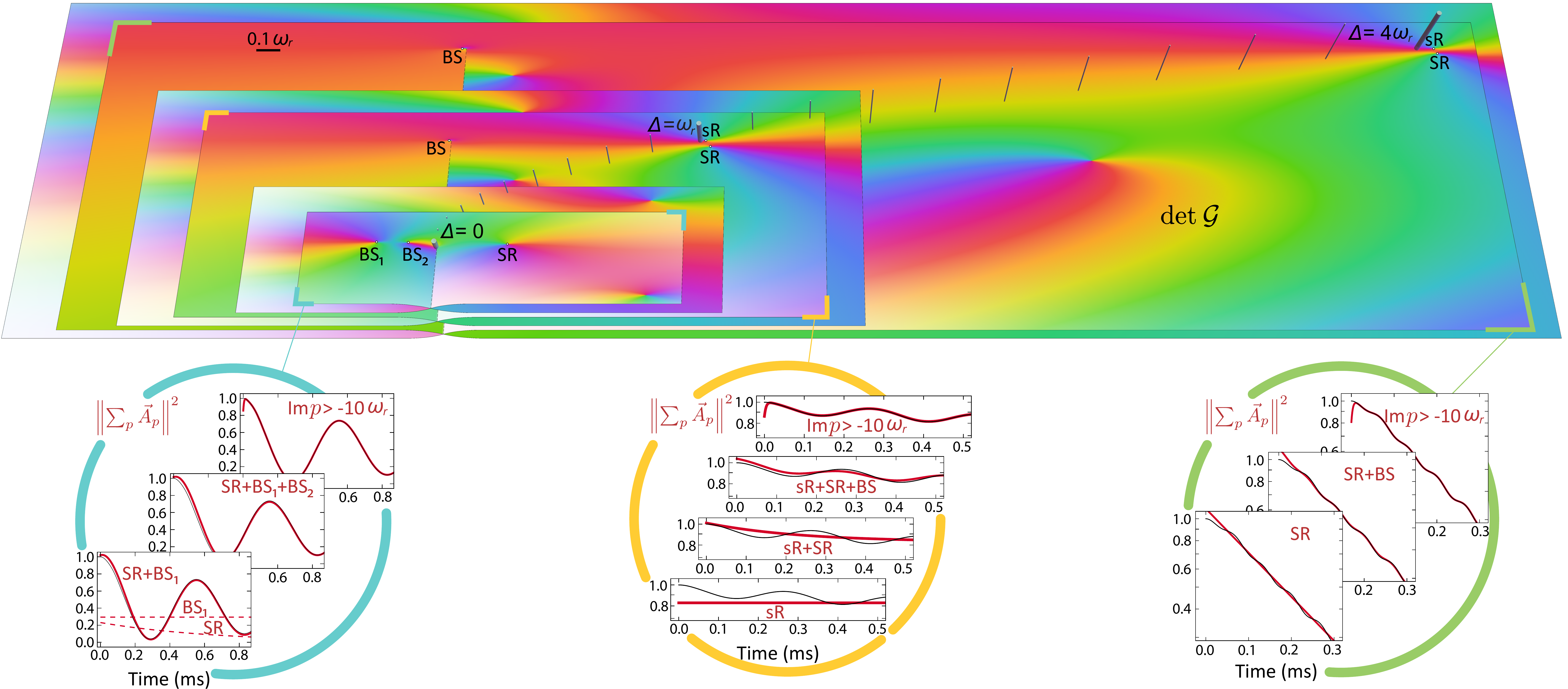}
    \caption{\textbf{Spectral contributions to the simulated dynamics of 3 emitters.} The domain colorings of $\det\matrixG$ are represented on Riemann surfaces, along with their analytic extensions, for $s_z=8$ and $(\Delta,\Omega)/\omega_\textsf{r}=(0,0.6)$, (1,0.42) and (4,1) from left to right (the vertical bars represent the dispersion $k(\Delta)$, to which the origin of the imaginary axis for each $\Delta$ is aligned).  The zeroes and branch cut of this function define the decay dynamics of the emitters, which are presented in the bottom panels. The black lines represent the simulated dynamics, while the red lines account for the various spectral contributions; with the main ones coming from superradiant (SR), subradiant (sR) and bound states (BS).
    }
\label{FIGPoles}
\end{figure*}

\subsection*{\label{APP:Analytic} Spectral structure and bound states in the continuum (BIC)}

We use the analytic formalism developed in \cite{Lanuza2022} to identify the spectral contributions including those of bound states for the initial $q=0$ timed-Dicke state in an array of 3 sites. The excitation amplitudes of each site can be broken down as $(A_1,A_2,A_3)^\intercal=I_1(t)/\sqrt{3}(1,1,1)^\intercal+I_2(t)/\sqrt{3}(0,1,0)^\intercal$ since the even parity of the initial state ($I_1(0)=1, I_2(0)=0$) is preserved throughout the decay. The exact time evolution is given by the integrals
\begin{equation}
     I_j(t)=\frac{(-1)^j}{2\pi i}\int_{-\infty+i0^+}^{+\infty+i0^+}e^{-i\omega t} \matrixG_{2,3-j} {\det}^{-1}\matrixG\dif\omega,
\end{equation}
where
\begin{equation}
\begingroup
  \small   
  \thinmuskip=\muexpr\thinmuskip*1/8\relax
  \medmuskip=\muexpr\medmuskip*1/8\relax  
\begin{aligned}
\matrixG=(\omega-\Delta)\mathbf{Id}_2+i\left(\begin{array}{cc}
        \tilde G_0(\omega)+\tilde G_1 (\omega)+\tilde G_2 (\omega) & \tilde G_1 (\omega) \\
        \tilde G_1(\omega)-\tilde G_2(\omega) & \tilde G_0(\omega)-\tilde G_1(\omega)
    \end{array}\right)
    \end{aligned}
    \endgroup
\end{equation}
and
\begin{equation}
\begin{aligned}
    &\tilde{G}_n(\omega)=\frac{\Omega^2 \pi^{1/2}}{8\sqrt{\omega \omega_\textsf{r}\sqrt{s_z}}}e^{-\omega/(\omega_\textsf{r}\sqrt{s_z})}\times
    \\&
    \left[e^{i n\pi\sqrt{\omega/\omega_\textsf{r}}}\operatorname{erfc}\left(-i\sqrt{\frac{\omega}{\omega_\textsf{r}\sqrt{s_z}}}-\frac{n\pi\sqrt[4]{s_z}}{2}\right)+n\leftrightarrow -n \right]
\end{aligned}
\end{equation}
is the Fourier transform of the bath correlation function $G_{j,j+n}(\tau)$.

These integrals can be decomposed using the residue theorem into a non-Markovian decay branch-cut at the continuum edge (at $\omega=0$) and a set of infinite simple poles satisfying $\det\matrixG(\omega_p)=0$. Most of these have small contributions and decay extremely fast (well beyond superradiant rates), but they provide a mechanism to avoid the ``superradiant paradox" \cite{Sinha2020}, as they help the initial decay to match the dynamics of a single emitter (cf. Fig.~\ref{FIG3}a).

Apart from these, we identify 3 poles representing a bound state, a subradiant, and a superradiant state. Their amplitudes are given by
\begin{equation}
    \vec{A_p}=\frac{e^{-i\omega_p t}}{\sqrt{3}(\det\matrixG)'(\omega_p)}\left(\begin{array}{c}
\matrixG_{2,2}(\omega_p) \\
\matrixG_{2,2}(\omega_p)-\matrixG_{2,1}(\omega_p) \\
\matrixG_{2,2}(\omega_p)
  \end{array}\right).
\end{equation}
They are tabulated together with their corresponding frequencies in Table~\ref{TABLE1}, and their contributions to the decay dynamics are represented in Fig.~\ref{FIGPoles}. Notice that at $\Delta=4\omega_\textsf{r}$ the dynamics is dominated by the superradiant state but at $\Delta=\omega_\textsf{r}$ besides the subradiant state, the superradiant state has a non-negligible contribution. We note that the subradiant state has an extremely small imaginary part, so technically it is a quasi-bound state in the continuum, but its decay is too small for us to resolve experimentally and thus we treat it as a BIC in the main text. The fact that our quantum emitters have a finite size prevents this state from being perfectly bound and from having the ideal amplitude $\vec{A}\propto (1,2,1)^\intercal$, which otherwise allows for perfectly destructive interference between the three point-like emitters ($\sum_j A_{j}e^{ik_\textsf{r} dj}=0$). The matter-wave component $B_k$ of the BIC follows from the $A_j$ as
\begin{equation}
\label{eq:Bk}
    B_{k} = \frac{\Omega {\textstyle\sum}_j \gamma_{k,j} A_{j}}{2(\omega_{\text{sR}} - \omega_k)}.
\end{equation}
We use this to compute the momentum and position distributions $n_b(k) = |B_k|^2$ and $n_b(z) = |B_{z}|^2$ of the BIC in the lineout plots of Fig.~\ref{FIG3}c (dashed lines), where $B_{z} \propto \sum_k e^{i k z} B_k$.

A heuristic fit of the observed population dynamics at $\Delta = \omega_\textsf{r}$ in Fig.~\ref{FIG3}a (ii) with an equation motivated by the analysis, $|\alpha_t e^{-i\omega t} + (1 - \alpha_0)|^2$, where $\alpha_t = \alpha_\infty + (\alpha_{0} - \alpha_\infty) e^{-3\Gamma_1 t/2}$, returns initial and asymptotic population averages, $|\alpha_0|^2 = 0.93(1)$ and $|\alpha_\infty|^2 = 0.84(2)$. These results qualitatively agree with the initial joint contribution from the super- and subradiant states (BIC), $||\vec{A}_{\text{sR},0}+\vec{A}_{\text{SR},0}||^2 = 1.01(2)$, and the asymptotic contribution from the subradiant state only, $||\vec{A}_{\text{sR},0}||^2 = 0.83(1)$ (cf. Table. \ref{TABLE1}). The beat frequency from the fit, $\omega = 0.95(2) \times \omega_\textsf{r}$, is also comparable to the frequencies of the two states, approximately at $\omega_\textsf{r}$.

Finally, at $\Delta=0$, there are two bound states below the edge. This feature does not appear with one or two symmetric emitters, and would ideally lead to persistent oscillations of very low frequency. They correspond to a redistribution of the population of the emitters from the center to the sides and back, in some analogy to vibrations in a molecule that are of low frequency compared to electronic excitations.\\

\begin{figure}[!htbp]
\centering
    \includegraphics[width=.99\columnwidth]{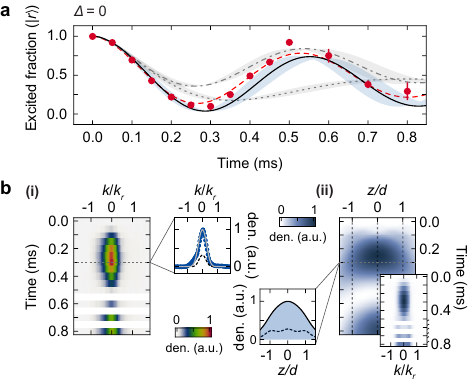}
    \caption{\textbf{Collective dynamics at the continuum edge in the superfluid regime.} \textbf{a}, Excited fraction as a function of time for $(s_z,s_\perp)=(8,8)$, $\phi=0$ 
    with strong coupling $\Omega = 0.60(4)\omega_\textsf{r}$ at $\Delta = 0$ (red points).
    The solid (dot-dashed) line is our simulations of a 3-site array with an excitation spread over the array (localized in a middle site); the dotted line is that of an isolated site; shaded areas represent the uncertainty in $\Omega$. The red dashed line is a fit to the beating of a dissipative and a bound state with the decay rate fixed by our analytic theory.
    \textbf{b}, (i) Momentum distribution of matter waves versus time. The lineout plot shows the data at $0.3$~ms (blue points) along with our simulation and bound-state contributions from our analytic theory (gray solid and dashed lines). (ii) Simulated position and momentum distributions of matter waves versus time, with a lineout plot at $0.3$~ms. The dashed vertical lines are the positions of the emitters.
    All data are averages of at least $3$ measurements; the error bars show the standard error of the mean.
    }
\label{FIGD0}
\end{figure}

\subsection*{\label{APP:CollectiveD0} Collective dynamics at the continuum edge}

As mentioned in the main text, the diverging mode density near $k=0$, i.e. $\rho(E) \propto 1/v_g \rightarrow \infty$, can also cause strong non-Markovian behavior \cite{Stewart2017,Lanuza2022}. We examine this regime by tuning our emitters to $\Delta=0$ and measure the population of the excited state prepared in a TDS with $\phi=0$ (Fig. \ref{FIGD0}a).
An oscillation with a near-maximum visibility is observed, much stronger than that expected for both an isolated emitter (which shows fractional decay with damped, weak oscillations) and for an exited emitter surrounded by empty neighbors (cf. \cite{Krinner2018,Stewart2020}) Instead, the  behavior resembles that of a TDS with $\phi=0$ (Fig. \ref{FIGD0}a solid line) strongly coupled to the $k = 0$ mode. This is confirmed in the momentum distribution of matter waves, which is concentrated near $k=0$, in agreement with our finite-array simulation (Fig. \ref{FIGD0}b (i); while $v_g=0$ here, our simulation indicates that the matter waves can still disperse toward neighboring emitters). In position space, this corresponds to the radiation trapped inside the array including contributions from bound states, with very little escaping to the outside (Fig. \ref{FIGD0}b (ii)).

Motivated by our analytic theory, we fit the data to the beating between a dissipative and a bound state $|\alpha_0 e^{-i\omega t - \Gamma_{\text{SR}} t/2} + (1-\alpha_0)|^2$, where $\Gamma_{\text{SR}} = 2\text{Im} (\omega_{\text{SR}}) \approx 2\pi \times 0.24$~kHz (Table \ref{TABLE1}), yielding a frequency $\omega = 2\pi \times 1.86(2)$~kHz and a bound-state population $|1-\alpha_0|^2 = 0.43(2)$ in a qualitative agreement with the frequency difference $\omega_{\text{SR}} - \omega_{\text{BS}_1} = 2\pi \times 1.84(11)$~kHz between the dominant states and the joint contribution of bound states $|| \vec{A}_{\text{BS}_1,0} + \vec{A}_{\text{BS}_2,0} ||^2 = 0.30(1)$. We note that the measured frequency is very close to the single-emitter coupling $\Omega = 0.6 \omega_{\textsf{r}} \approx 2.2$~kHz corrected by the Franck-Condon overlap normalized to a single Wigner-Seitz cell ($\approx 0.82$). This reflects the fact that in the limit of large coherence length, the dynamics reduces to the Rabi oscillation between two quasimomentum eigenstates due to the translational symmetry \cite{Lanuza2022}, as also observed in a diffraction experiment of condensates from a state-selective optical lattice \cite{Reeves2015}.

\begin{figure}[!htbp]
\centering
    \includegraphics[width=0.99\columnwidth]{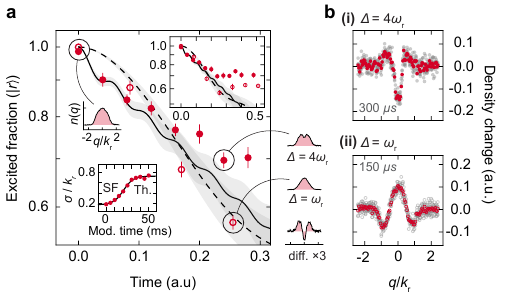}
    \caption{\textbf{Radiative decay of thermal excitations.} \textbf{a,} Excited fraction for two detunings $\Delta = 4\omega_\textsf{r}$ and $\omega_\textsf{r}$ (red points and circles), with coupling strength set to $\Omega/\omega_\textsf{r} = 1.00(7)$ and $0.60(4)$. A thermal gas of $\ket{r}$ atoms is prepared in shallow lattices, $(s_z,s_\perp)=(8,8)$, by heating a condensate via intensity modulation with an average amplitude $\approx 30$\% and a frequency $500$~Hz for a duration of $40$~ms. The solid and dashed lines are simulations of single-emitter decay for the same parameters. To make the time scale comparable to that of Fig.~\ref{FIG4}a, for $\Delta=4\omega_\textsf{r} (\omega_\textsf{r})$, the time axis (in ms) is rescaled by the ratio $\Gamma_1/\Gamma_{1}'=0.8(1.7)$ where $\Gamma_1$ is the Markovian single-emitter decay rate for the current parameters and $\Gamma_{1}'$ is that of Fig.~\ref{FIG4}a. For $\Delta=4\omega_\textsf{r} (\omega_\textsf{r})$, quasimomentum distributions (q.d.) of $\ket{r}$ atoms are shown at $t=0$ and $300$~$\mu$s ($150$~$\mu$s). The top-right inset shows the same data including longer times. Also shown is the heating of atoms versus the lattice modulation time characterized by the width of momentum peak \cite{Kwon2022}; the solid line is a sigmoidal fit.
    \textbf{b,} (i) Change in the normalized q.d. of $\ket r$ atoms after $300$~$\mu$s for $\Delta = 4\omega_\textsf{r}$. (ii) Same but after $150$~$\mu$s for $\Delta = \omega_\textsf{r}$.
    All data are averages of at least $3$ measurements with the error bars from the SEM (gray points and circles are raw data).
    }
\label{FIGTh}
\end{figure}

\subsection*{\label{APP:Thermal} Radiative decay of thermal excitations}

Here we present the decay data for thermal atoms in a weakly interacting regime. Such a system is prepared by periodically modulating the lattices in order to slightly heat up the $\ket r$ atoms before applying the coupling pulse (see Fig. \ref{FIGTh}a). 
For both $\Delta = 4\omega_\textsf{r}$ and $\omega_\textsf{r}$, the decay of the excited fraction resembles that of isolated emitters in contrast to the behavior of a superfluid, and stalls after a certain time $\sim 0.2$~ms. To understand this behavior, we consider a thermal ensemble of $\ket r$ atoms that are spread over all available quasimomenta $q$ in the first Brillouin zone. Since only those near the superradiant point $q = \tilde{k}(\Delta)$ can be emitted ($\tilde{k}$ being the matter-wave quasimomentum as defined in the main text), the decay stops once only the subradiant states are left. Consequently, looking at the quasimomentum distribution of $\ket r$ atoms over time, we find that the populations at the emitted quasimomenta are depleted, e.g. at $q = 0k_\textsf{r}(\pm1k_\textsf{r})$ for $\Delta = 4\omega_\textsf{r}(\omega_\textsf{r})$ as shown in Fig. \ref{FIGTh}b.
We note that these intuitive results are the exact opposites of those seen in Fig.~\ref{FIG4}c in the Mott regime, where the gains rather than losses appear at the same $q$.

\paragraph*{\bf Acknowledgements} 
We thank J. Kwon and H. Huang for experimental assistance, M. G. Cohen for discussions, and J. Kwon and M. G. Cohen for a critical reading of the manuscript.
This work was supported by NSF PHY-1912546 and PHY-2208050. 
\paragraph*{\bf Author contributions} Y.K., A.L. and D.S. conceived the experiments. Y.K. took the measurements and analysed the data. Numerical simulations and analytical descriptions were developed by Y.K. and A.L., respectively. The results were discussed and interpreted by all authors. D.S. supervised the project. The manuscript was written by Y.K. and D.S. with critical contributions from A.L.

\end{document}